\documentclass{llncs}
\usepackage{cite}
\usepackage{amsmath,amssymb,amsfonts}
\usepackage{algorithmic}
\usepackage{graphicx}
\usepackage{textcomp}
\usepackage{booktabs}
\usepackage{url}
\usepackage{csquotes}
\usepackage{amssymb}
\def\BibTeX{{\rm B\kern-.05em{\sc i\kern-.025em b}\kern-.08em
    T\kern-.1667em\lower.7ex\hbox{E}\kern-.125emX}}

\begin{document}

\title{ProML: A Decentralised Platform for Provenance Management of Machine Learning Software Systems}
\titlerunning{ProML}

\author{Nguyen Khoi Tran\inst{1} \and 
Bushra Sabir\inst{1} \and 
M. Ali Babar\inst{1} \and 
Nini Cui\inst{1} \and 
Mehran Abolhasan\inst{2} \and 
Justin Lipman\inst{2}}

\institute{The University of Adelaide, Adelaide, Australia \and
University of Technology Sydney, Sydney, Australia}

\maketitle

\begin{abstract}
    Large-scale Machine Learning (ML) based Software Systems are increasingly developed by distributed teams situated in different trust domains. Insider threats can launch attacks from any domain to compromise ML assets (models and datasets). Therefore, practitioners require information about how and by whom ML assets were developed to assess their quality attributes such as security, safety, and fairness. Unfortunately, it is challenging for ML teams to access and reconstruct such historical information of ML assets (ML provenance) because it is generally fragmented across distributed ML teams and threatened by the same adversaries that attack ML assets. This paper proposes ProML, a decentralised platform that leverages blockchain and smart contracts to empower distributed ML teams to jointly manage a single source of truth about circulated ML assets' provenance without relying on a third party, which is vulnerable to insider threats and presents a single point of failure. We propose a novel architectural approach called Artefact-as-a-State-Machine to leverage blockchain transactions and smart contracts for managing ML provenance information and introduce a user-driven provenance capturing mechanism to integrate existing scripts and tools to ProML without compromising participants' control over their assets and toolchains. We evaluate the performance and overheads of ProML by benchmarking a proof-of-concept system on a global blockchain. Furthermore, we assessed ProML's security against a threat model of a distributed ML workflow.
\end{abstract}


\section{Introduction}

Large-scale Machine Learning (ML) based Software Systems are increasingly developed by distributed interdisciplinary teams \cite{nahar2022collaboration}. For instance, \texttt{booking.com}, one of the largest online travel agencies, employs \textit{dozens} of multi-functional teams consisting of software developers, user interface designers, ML researchers, and ML engineers to develop and operate ML models \cite{bernardi2019150}. Due to the emerging trend of outsourcing and crowdsourcing in ML engineering \cite{bernardi2019150,nahar2022collaboration}, ML teams might not even reside in the same organisation. Let us consider the following running example. A startup A has an ML application idea but lacks the ML skillset and infrastructure to realise it. Therefore, A outsources data collection and model development to companies B and C. The company B acts as a \textit{dataset administrator}, collecting data from relevant sources (social media, open-source intelligence, official statistics) and labelling the data by crowdsourcing via a marketplace such as Amazon Mechanical Turk, a common practice in ML engineering \cite{paolacci2010running}). Model development happens at C, where various \textit{model developers} carry out workflow activities such as preprocessing data, engineering features, training and testing models. The trained models in the form of binaries or neural net architecture and weight values are returned to the \textit{operators} and \textit{auditors} at A for testing in a beta product. 

Insider threats can compromise ML assets (models and datasets) at many points in the described ML workflow. For instance, annotators in the crowdsourcing platform can be bribed to mislabel data samples to enable poisoning attacks \cite{he2020towards}. Some members of the model development team at C might be bribed to swap the trained model with a poisoned one before delivery. Following problem reports from A, organisation C might detect the model swapping following an internal investigation. However, they might not disclose the issue to protect their reputation, thus harming A. Therefore, all workflow participants, especially the auditors at A, require a complete history of ML assets \textit{(ML provenance)} that show how and by whom workflow activities were performed in order to evaluate various quality attributes of ML assets, such as security, safety, and fairness. Unfortunately, \textit{ML provenance information is generally fragmented} across a distributed ML workflow because participants can only record the provenance related to their activities and store the information in private silos. Moreover, even if the auditors at A can retrieve and reconstruct a complete history of ML assets, \textit{such information might not be trustworthy} as the retrieved provenance records might also have been compromised by the insider threats that attacked ML assets. 

This paper proposes \textbf{ProML} (\underline{\textbf{Pro}}venance information of \underline{\textbf{M}}achine \underline{\textbf{L}}earning), a \textit{decentralised} platform that leverages blockchain and smart contracts to manage ML provenance in a distributed ML workflow. Unlike the existing centralised provenance management solutions (e.g., \cite{Schelter:2017:1,Souza:2020:1}), ProML does not require workflow participants to appoint a leader or hire a third party to operate it. Instead, the platform distributes the right and responsibility to access and manage ML provenance information across stakeholders of an ML workflow, such as organisations A, B, and C in the running example. Stakeholders join the platform by running identical software clients called \textit{ProML nodes} that maintain a private blockchain and run smart contracts to store and update ML provenance information. On-blockchain records serve as a \textit{single source of truth} about the history of the circulated ML assets, addressing the fragmentation of ML provenance. On-blockchain software programs called smart contracts implement the necessary computation to process provenance updates submitted by workflow participants. Leveraging blockchain and smart contracts brings about various advantages, including data immutability, non-repudiation, integrity, transparency, and fairness \cite{Wood:2014:1,Xu:2018:2,Tran:2020:1,Cachin:2017:1}. It should be noted that ProML directly involves stakeholders in blockchain operations (via ProML nodes) rather than outsourcing the responsibilities to a remote blockchain, an approach taken by many existing blockchain-based provenance systems for ML models \cite{Stokes:2021:1} and datasets \cite{Sutton:2017:1,Neisse:2017:1,Baracaldo:2017:1,Demichev:2018:1,Zhang:2018:1,Kumar:2018:1,Dang:2020:1,Moller:2021:1,Khatal:2021:1}. The direct participation provides stakeholders with complete control and access to the provenance information, reduces costly transaction fees and privacy risks associated with remote blockchains. We propose a novel architectural approach called \textit{Artefact-as-a-State-Machine (ASM)} to utilise blockchain transactions and smart contracts for storing and updating ML provenance information.

Rather than implicitly monitoring and extracting ML provenance from workstations or big data processing clusters, we propose a \textit{user-driven provenance capturing process} to help participants control what and how ML provenance is extracted without exposing them to the underlying complexities. In particular, every ProML node offers services and APIs for submitting provenance updates. Organisation C can embed services offered by ProML into their existing ML training scripts or provenance tools such as MLFlow \footnote{(mlflow.org)} and Sacred \footnote{(Github/IDSIA)} to control the reported information, preventing issues such as Data Use Agreement (DUA) violations. A ProML node belonging to C is considered trusted within C and can only be used by C to access and submit ML provenance. 

We evaluate ProML's performance and overheads by deploying and benchmarking a proof-of-concept system on a global test blockchain network called Ropsten\footnote{\url{https://ropsten.etherscan.io}} to simulate a globally distributed ML workflow. Furthermore, we analysed the security of ProML against a threat model of a distributed ML workflow. On the Ropsten blockchain network, we found that the framework takes around 16 seconds to capture a provenance update and 2.5 minutes to finalise it with high confidence. These figures are negligible to the overall timeframe of a distributed ML workflow, which generally takes at least seven months to bring ML model ideas to production \cite{dotscience:2019:1} and at least eight days to deploy a trained model \cite{Algorithmia:2020:1}. Overhead-wise, according to the conversion rate obtained in April 2022, we found that registering a new asset with ProML costs around USD $\$160$ and submitting a provenance record costs around USD $\$47$. It should be noted that these monetary costs do not apply to private blockchains that ProML targets because participants in a private blockchain can generate an arbitrary amount of cryptocurrency to fund their provenance submission. 

The major contributions of this paper include:
\begin{itemize}
    \item Conceptualisation and architecture of a decentralised platform based on blockchain and smart contracts for secure management of ML provenance in a distributed ML workflow. 
    \item A novel architectural approach for leveraging blockchain transactions and smart contracts to securely store and update ML provenance information in a distributed ML workflow.
    \item A user-driven approach to capture ML provenance information that preserves users' control, eases the integration with users' existing scripts and toolchains and abstracts the complexity of blockchain interaction. 
\end{itemize}

\section{Preliminary}
\label{sec:Preliminary}

A \textit{blockchain} is a shared-write database and a secure distributed computer operated by mutually distrusting parties \cite{Cachin:2017:1}. Formally, it is a transaction-based state machine (Eq. \ref{eq:blockchain_state_machine}) replicated across blockchain participants \cite{Wood:2014:1,Tran:2020:1}. 

\begin{equation}
    \sigma_{t+1}\equiv \Upsilon(\sigma_t, T) \label{eq:blockchain_state_machine}
\end{equation}

State transitions are initialised and controlled by \textit{transactions} $T$, which are instructions coming from \textit{blockchain accounts}. These instructions convey diverse messages, from moving fund between accounts to invoking user-defined software programs, also known as \textit{smart contracts}. Transactions are digitally signed by senders to prevent forgery and tamper as they are distributed amongst blockchain participants. The signatures also prevent senders from falsely denying their transactions. 

Blockchain participants calculate the new state $\sigma_{t+1}$ independently using the deterministic state transition function $\Upsilon$. Using a fault-tolerant consensus protocol such as Proof-of-Work (PoW), Proof-of-Authority (PoA), Proof-of-Stake (PoS), and Practical Byzantine Fault Tolerance (pBFT), participants reach an agreement on $\sigma_{t+1}$. The transaction leading to the accepted $\sigma_{t+1}$ is added to an ordered transaction list called a \textit{ledger}. All participants hold a copy of the ledger, which represents the canonical history of a blockchain. The ledger is usually stored in a \textit{secure data structure called blockchain}, which uses cryptographic hashes to prevent and detect tampering of content and order of transactions. 

Some blockchain protocols such as Ethereum allows users to embed arbitrary software programs called smart contracts to $\Upsilon$. Smart contracts are generally as stateful objects that has their own blockchain address, internal variables, and functions to act upon those variables.

\section{ProML Platform}
\label{sec:framework}

\begin{figure}[ht]
    \centering
    \includegraphics[width=1\textwidth,,trim={0cm 9.5cm 2cm 0cm},clip]{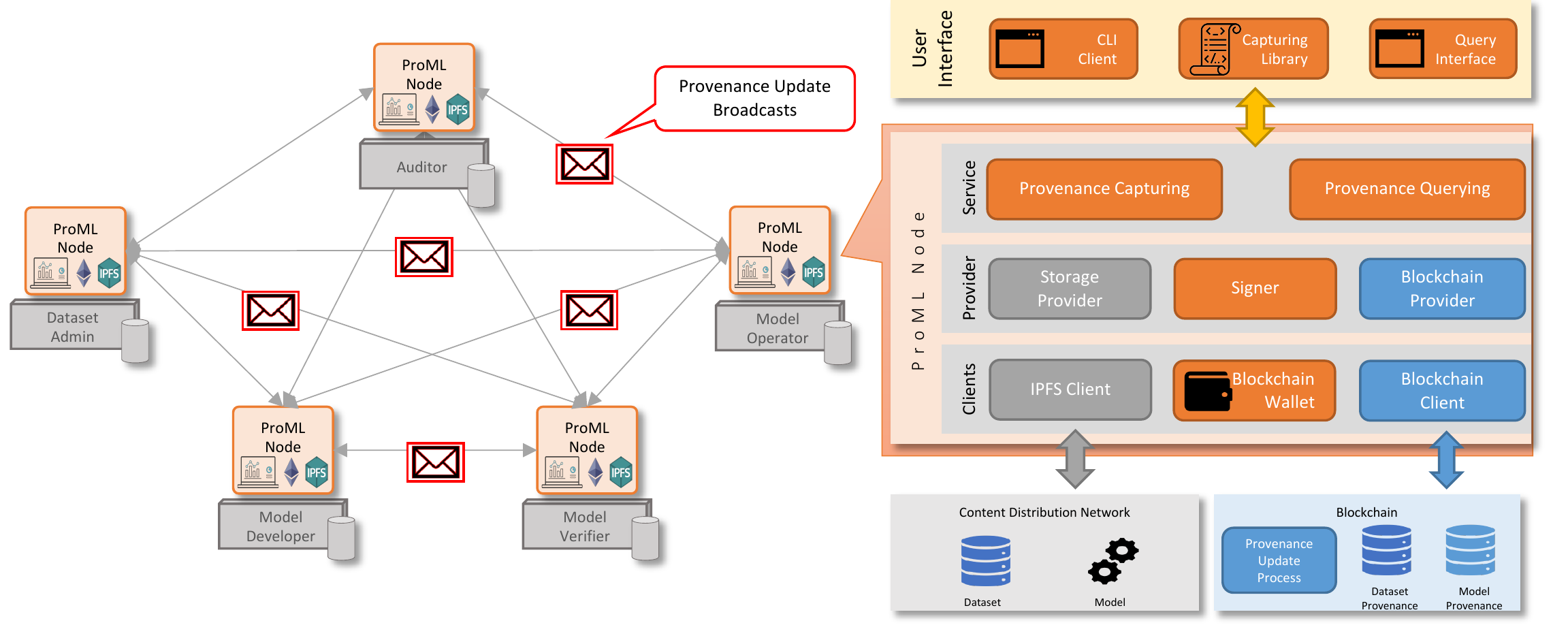}
    \caption{ProML Platform for a distributed ML workflow with five participants}
    \label{fig:ProML_deployment_arch}
\end{figure}

Every deployment of the ProML platform consists of multiple identical software clients called ProML nodes, deployed and operated independently by workflow participants. The admission, authorisation, and governance of participants are beyond the scope of our paper. We assume that such decisions are in place before ProML is deployed. Figure \ref{fig:ProML_deployment_arch} depicts a ProML deployment for five participants. 

Components within a ProML node are structured into three layers. The \textit{client} layer contains software clients implementing blockchain and other distributed protocols used by ProML. This layer also contains a \textit{blockchain wallet} that holds a participant's private keys to sign blockchain transactions on their behalf. Upon deployment, clients from different ProML nodes connect to form a blockchain and other decentralised infrastructure used by ProML, such as a private content distribution network based on the Interplanetary File System (IPFS\footnote{\url{https://ipfs.io}}). ProML deploys smart contracts on the deployed blockchain to capture and update ML provenance information. Section \ref{sec:artefact-as-a-state-machine} presents our proposed architectural approach called Artefact-as-a-State-Machine (ASM) for engineering these contracts. 

The \textit{provider} layer contains service wrappers that abstract the technical complexities and provide a consistent Application Programming Interface (API) to the clients. For instance, the \textit{blockchain provider} can offer a consistent API with high-level services such as ``send a transaction'' and transform the function calls into proper messages expected by an underlying blockchain client. \textit{Storage provider} and \textit{signer} perform similar tasks for storage clients (e.g., IFPS) and wallets. 

The service layer contains high-level services for accessing and updating ML provenance. Participants can invoke these services ad-hoc via the provided software clients or programmatically using a software library. Section \ref{sec:semi_automatic_provenance_capture} presents our proposed user-driven approach the leverages these components for capturing ML provenance. 

\subsection{User-Driven Provenance Capture}
\label{sec:semi_automatic_provenance_capture}

The provenance capturing process is a collaboration between participants and their trusted ProML nodes. Participants \textit{initiate the process} and supply the necessary provenance records ($pm_i$ or $Prov_{DS_i}$). ProML nodes \textit{finish the process} by crafting and and submitting the corresponding blockchain transactions ($tx_{pm_i}$ or $tx_{Prov_{DS_i}}$) on behalf of the participants with their blockchain credentials. 

\begin{figure}[ht]
    \centering
    \includegraphics[width=0.95\textwidth,,trim={0.2cm 7.8cm 3cm 0cm},clip]{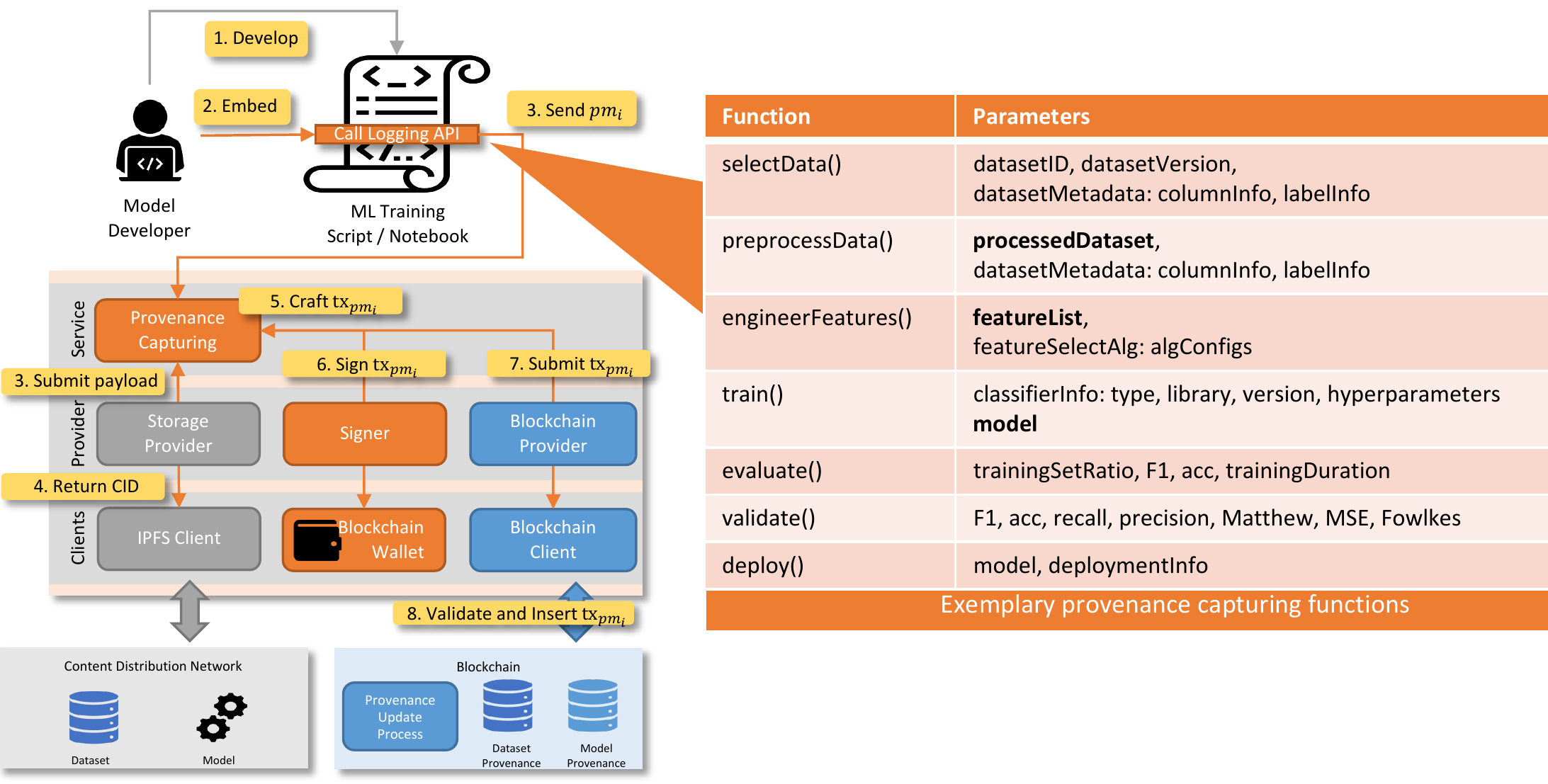}
    \caption{User-Driven Provenance Capturing Mechanism}
    \label{fig:semi-automated-capture}
\end{figure}

Figure \ref{fig:semi-automated-capture} depicts the process for capturing and submitting ML provenance from a training script. Participants can submit $pm_i$ or $Prov_{DS_i}$ to ProML in an ad-hoc manner via a commandline interface (CLI) client, programmatically via a software library, or directly by invoking the provenance capturing service of of their trusted ProML node. This procedure has the following advantages: 
\begin{itemize}
    \item \textit{Transparency and control:} ML development activities can utilise and produce sensitive data, which must conform to predefined Data Use Agreements (DUAs). Therefore, users require visibility and control over the information captured and propagated by the provenance management system so that they can verify and therefore trust the system. The user-driven procedure of ProML meets this ends. 
    \item \textit{Ease of Integration:} By offering software libraries and service interfaces, ProML allows users to integrate the provenance capturing mechanism into their toolchains in a way that is suited to their operating situation and setup. 
    \item \textit{Simplicity:} Having ProML nodes handle the complexities of interacting with blockchain helps to simplify the usage and avoid security mistakes.
\end{itemize}

The following section presents how different components in a ProML node work to capture some prominent life cycle events of ML assets. 

\noindent\textbf{Registering a model or dataset:} Asset registration creates an on-blockchain representative of an ML asset and begins its life cycle with ProML. When registering a dataset, ProML also requires the dataset's metadata, address, and the identifier of its ancestor if available. The \textit{provenance capturing service} uses the provided information ($pm_i$ or $Prov_{DS_i}$) to construct blockchain transactions for deploying smart contracts that represent the given models or datasets. It uses the \textit{signer} service, which holds participants' private keys, to sign the constructed transactions with participants' credentials. Following the signing, the provenance capturing service invokes the \textit{provider} service, which wraps around a blockchain client, to publish the transactions. The registration completes when the transactions have been added to the ledger, making the smart contracts available on the blockchain. ProML returns the addresses of the deployed contracts as asset identifiers. Details of the involved smart contracts would be elaborated in Section \ref{sec:artefact-as-a-state-machine}.

\noindent\textbf{Recording model's life cycle events:} ProML provides model developers with a software library containing functions that triggers the reporting of different actions performed on a model (selecting and preprocessing data, engineering features, training, evaluating, validating, and deploying model). The parameters passed to these functions are $payload$ that describe inputs $in$, outputs $out$, and parameters $param$ of actions. The number and types of functions making up the library closely relate to the on-blockchain representatives of ML models and reflect an agreed ML workflow amongst participants. By controlling whether to call these functions and what to provide, participants control the provenance information leaving their trust domain. Figure \ref{fig:semi-automated-capture} presents the process and exemplary provenance capturing functions. 

When the embedded functions are executed, they submit provenance records $pm_i$ to a trusted ProML node, whose \textit{provenance capturing service} constructs transactions $tx_{pm_i}$ that address the smart contract representing the ML models being recorded, and submits the transactions with the help of the \textit{signer} and \textit{provider} services. 

\noindent\textbf{Publishing a model or dataset:} The provenance records of a model or a dataset publication contain large payloads such as CSV files, databases, executable binaries, or neural network architectures and weight values. Storing such payloads on-chain is discouraged to prevent scalability issues of the distributed ledger. ProML leverages the \textit{off-chain data storage design pattern} \cite{Xu:2018:2} to secure the payloads associated with provenance records without keeping those payloads on the blockchain. Specifically, the provenance capturing service can offload the payloads to a distributed content-addressable storage, such as a private IPFS cluster, and keep the pointers to those payloads in the on-chain provenance records. The \textit{storage provider} service is responsible for managing and interacting with off-chain storage such as IPFS. 

\subsection{Artefact-as-a-State-Machine}
\label{sec:artefact-as-a-state-machine}

Mapping information and processes onto blockchain constructs is a prerequisite for storing and managing them on a blockchain. We propose an architectural approach to guide this mapping based on the following observations:
\begin{enumerate}
    \item \textbf{Every ML asset has a set of states according to its workflow.} For example, an ML model reaches the ``trained'' state after being fitted to a dataset by a training algorithm. 
    \item \textbf{Actions performed on an ML asset can trigger a state transition.} For instance, the training activity pushes an ML model from a previous stage (engineered features) to the next stage (trained). 
    \item \textbf{ML provenance is the record of a state transition and the responsible party.} In other words, it describes the input, output, parameters of a conducted activity and the identifier of the participant ($P_{id}$) carrying it out. It should be noted that provenance records are valid and valuable even if the reported activities failed to update the state of an asset. These records show that a participant has conducted a wrong action on it at a certain time.
\end{enumerate}

\begin{figure}[ht]
    \centering
    \includegraphics[width=1\textwidth,,trim={1cm 0cm 2cm 0cm},clip]{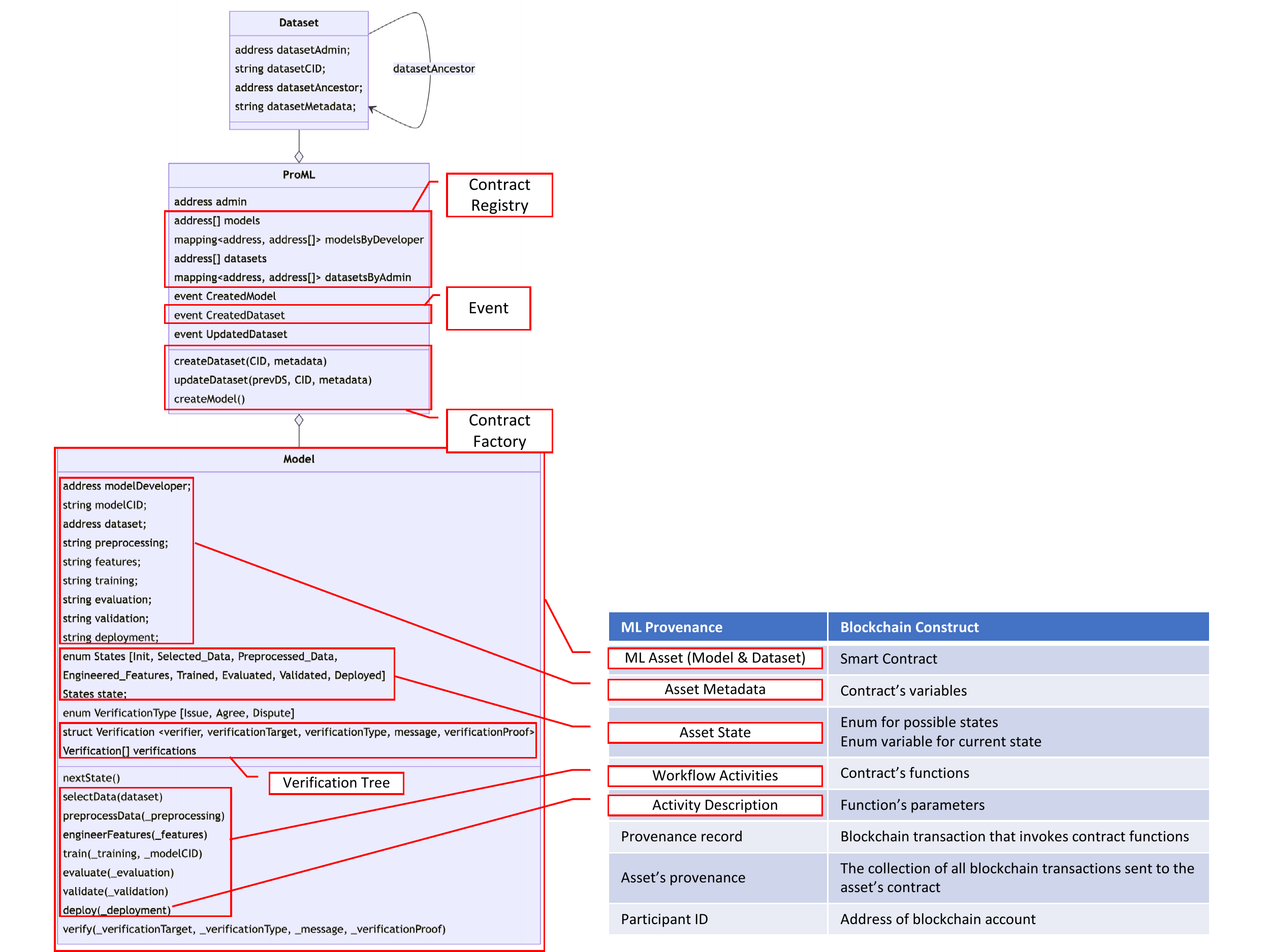}
    \caption{Exemplary smart contract implementation of ASM}
    \label{fig:ProML_class_diagram}
\end{figure}

The above observations suggest that ML assets can be modelled as state machines from a provenance perspective. Mapping this state machine representation of ML assets to blockchain constructs is straightforward, as blockchain by design is a replicated state machine (Section \ref{sec:Preliminary}). ML models and datasets map to smart contract instances because they are individually addressable blockchain objects that carry internal variables and functions to act upon those variables. The address of a smart contract becomes an asset's identifier. Internal variables of a smart contract can be used to store an asset's state and metadata. Functions of a smart contract represent workflow activities that update an asset's state. Function parameters can be used to capture the $payload$ of a provenance record corresponding to a workflow activity. Participants record a provenance update $pm_i$ by submitting a blockchain transaction to the smart contract to invoke a function corresponding to the reported workflow activity. These transactions serve as provenance records. We call this approach of modelling ML assets as state machines to map them to blockchain constructs for management as \textit{Artefact-as-a-State-Machine, or ASM}. Figure \ref{fig:ProML_class_diagram} depicts an exemplary implementation of ASM with smart contracts written in Solidity programming language\footnote{\url{https://soliditylang.org}}. 


\noindent\textbf{Implementation challenges and Solutions:} Implementing ASM introduce three challenges listed below. We address these challenges by leveraging the existing design patterns for blockchain-based application \cite{Xu:2018:2}. A utility smart contract named \texttt{ProML} was introduced to implement these patterns. 
\begin{itemize}
    \item \textit{Providing a lightweight communication mechanism to notify participants of on-blockchain events such as model publication.} We leverage the embedded event functionality of many blockchain platform to emit events to off-blockchain software via the log portion of a ledger. Events are defined and emitted by the utility contract \texttt{ProML}.
    \item \textit{Creating and deploying smart contract instances from predefined templates.} We apply the factory contract design pattern by introducing factory functions in the \texttt{ProML} contract, which creates and deploys contract instances based on the templates held by \texttt{ProML}. 
    \item \textit{Maintaining pointers to the registered models and datasets.} We apply the contract registry pattern by adding into \texttt{ProML} arrays of addresses pointing to asset contracts as well as mappings between participants and the registered assets for looking up. 
\end{itemize}

\section{Performance and Cost Evaluation}
\label{sec:evaluation}

This section presents an evaluation of ProML's performance and operating costs, which impact the platform's feasibility in real-world scenarios. We consider performance as how fast ProML processes and appends provenance updates. The cost aspect indicates the necessary resources for workflow participants to submit ML provenance updates. 

\begin{figure}[ht]
 \centering
 \includegraphics[width=\textwidth,,trim={0cm 13cm 0cm 0cm},clip]{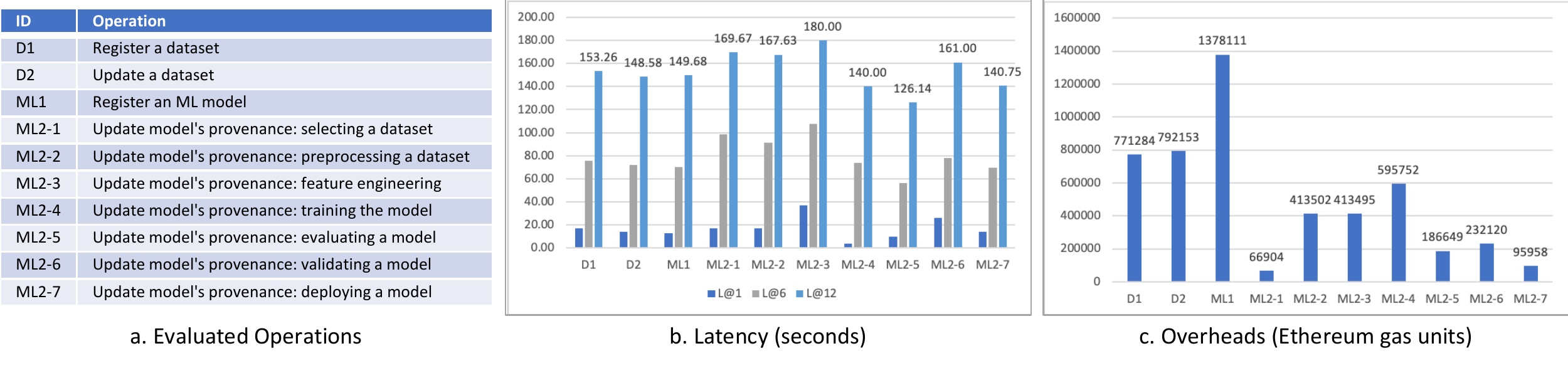}
 \caption{Latency and overheads of the evaluated operations}
 \label{fig:benchmark_results}
 \end{figure}

\subsection{Experimental Design}

We evaluated the performance and operating cost of ProML by benchmarking a proof-of-concept (PoC) implementation. The PoC leverages the Solidity programming language\footnote{\url{https://soliditylang.org}} to implement smart contracts according to the ASM approach. These smart contracts are compatible with many public and private blockchains that use Ethereum Virtual Machine (EVM) as the smart contract execution engine. The PoC's smart contracts were deployed on Ropsten, a global Ethereum blockchain network preserved for testing\footnote{\url{https://ropsten.etherscan.io}}. We chose this blockchain network due to its scale and similarity in configurations and performance characteristics with the Ethereum main network, which operates most high-profile and high-value blockchain-based software applications. We used \texttt{Ethers}\footnote{\url{https://docs.ethers.io/v5/}} for implementing the blockchain provider and signer services and \texttt{Node} for the provenance capturing and querying services.

\noindent\textbf{Procedure and Data:} The experimental workload consisted of ten operations that register and update a dataset and a model (Figure \ref{fig:benchmark_results}a). We triggered these operations sequentially with ten-second delays and tracked the submitted blockchain transactions to determine their latency and operating cost. This process was replicated ten times. We generated ML provenance for the experiment by training an intrusion detection classifier on the KDD-99 dataset\footnote{\url{http://kdd.ics.uci.edu/databases/kddcup99/kddcup99.html}} and utilising the provenance capturing procedure described in Section \ref{sec:semi_automatic_provenance_capture} to capture the data.

\noindent\textbf{Evaluation Metrics:} We measured the performance in terms of \textit{latency}, the amount of time between the submission of a provenance update ($tx_{Prov_{DS_i}}$ or $tx_{pm_i}$) and its inclusion in a blockchain. Every subsequent block represents an additional confirmation of the update, which strengthens its trustworthiness. We formalise a transaction's latency at the confirmation level $x$ as $L@x := t_0 - t_x$, where $t_0$ is the submission time, and $t_x$ is the time at which a transaction block representing the confirmation $x$ appears in the blockchain.

We measured the operating cost in terms of \textit{gas}, a unit that measures the computational effort required to process and store a transaction on an Ethereum blockchain, on which we deployed the proof-of-concept system. Transactions with higher gas values requires more computation and storage space on a blockchain than transactions with lower gas value. The Ethereum Yellow Paper \cite{Wood:2014:1} formally defines gas and a fee schedule. 

\subsection{Results}

\noindent\textbf{Latencies:}
Figure \ref{fig:benchmark_results}b presents the average latencies of various operations of ProML at one, six, and twelve confirmations. The framework takes around 16 seconds to record a provenance update to the Ropsten network ($L@1$). However, achieving confidence that a provenance update has been finalised requires around 2.5 minutes ($L@12$). 

The existing studies have shown that most companies take at least seven months to bring ML model ideas to production \cite{dotscience:2019:1} and at least eight days to deploy a trained model \cite{Algorithmia:2020:1}. Moreover, ML workflow activities generally do not need to wait for the provenance record of a previous step. Therefore, we argue that the provenance recording latency is negligible and would not impact the overall latency of the workflow.

\noindent\textbf{Operating Cost:}
Figure \ref{fig:benchmark_results}c compares the operating cost of the evaluation operations. The registration steps (D1, D2, and ML1) are the most costly operations because they involve creating and deploying additional smart contracts on the blockchain. Updating the provenance of an ML model (ML2-1 to ML2-7) incurs a lower cost, averaging at 280000 gas units. 

To put these figures in context, we convert them to the dollar value using the conversion rate obtained in April 2022 from the public Ethereum network (USD $\$ 0.000163542$ per gas unit\footnote{\url{https://etherscan.io/gastracker}}). According to the conversion rate, registering a new asset with ProML costs around USD $\$160$ and submitting a provenance record costs around USD $\$47$. It should be noted that these monetary costs do not apply to private blockchain networks where ProML aims to operate, because participants in a private blockchain can generate an arbitrary amount of cryptocurrency to fund their provenance submission.

\section{Security Evaluation}

This section evaluates the security of ProML. First, we define a threat model of a distributed ML workflow where ProML operates. Then, we present the security countermeasures provided by ProML and discuss their effectiness against the identified threats. 

\subsection{Threat Model}
\label{sec:threat_model}

ProML considers an adversary who aims to corrupt the provenance and verification of ML artefacts to cover up their tampering with a model \cite{Jentzsch:2019:1} or a dataset (e.g., poisoning attack \cite{Baracaldo:2017:1}). We assume that ML workflow participants are \textit{not anonymous} and possibly bound by legal contracts. We assume that the involved networks and infrastructures are protected by cloud service vendors or enterprise security mechanisms and thus safe from traditional threats, allowing us to focus on insider threats: authorised employees who have gone rogue or compromised by external attackers. 

Table \ref{tbl:ProML_threats} presents the threat model of ProML. Tampering threats denote unauthorised modification or destruction of artefacts and provenance information. Spoofing threats denote the forgery of provenance updates and verification results. Repudiation denotes that a participant falsely denies their previously provenance records to avoid responsibility. Finally, DoS threats denote attacks that corrupt datasets, models, and provenance information or disrupt the infrastructure necessary to access them. 

\begin{table}[hb]
    \centering
    \scriptsize
    \caption{Threat Model of ProML}
    \label{tbl:ProML_threats}
    \begin{tabular}{@{}lll@{}}
    \toprule
       & Threat      & Target                                                \\ \midrule
    T1 & Tampering   & At-rest data: Datasets, Models, Provenance Records    \\
    T2 & Tampering   & In-transit data: Datasets, Models, Provenance Records \\
    T3 & Spoofing    & Provenance records, verification results              \\
    T4 & Repudiation & Provenance records                                    \\
    T5 & DoS         & Data stores for datasets, models, provenance records  \\
    T6 & DoS         & Provenance capturing process                          \\
    T7 & DoS         & Provenance retrieval process                          \\ \bottomrule
    \end{tabular}
\end{table}


\noindent \textbf{Secury Countermeasures of ProML} ProML provides the following security countermeasures by leveraging blockchain, smart contracts, and a decentralised architecture. 

\begin{itemize}
    \item \textbf{Storing provenance records on a blockchain:} Blockchain data has tamper-resistance and high-availability by design \cite{Xu:2018:2,Tran:2020:1}. By storing provenance records on a blockchain, we protect them against unauthorised modification and destruction by insider threats. This countermeasure addresses threat T1, T2, and T5. 
    \item \textbf{Storing assets on a peer-to-peer content distribution network.} The storage facility of models and datasets presents a single point of failure if it is the only source of those assets. Therefore, ProML nodes form a private, peer-to-peer content network based on the IPFS protocol to offer participants an alternative resilient storage solution. This countermeasure addresses threat T5 and T7.
    \item \textbf{Anchoring off-blockchain artefacts to on-blockchain records using content identifiers.} ProML embeds a cryptographic representation (hash) of the registered models and datasets, allowing users to verify the integrity of any incoming asset. This countermeasure addresses threat T1 and T2.
    \item \textbf{Embedding the provenance recording process in smart contracts.} ProML implements the logic related to the registration and update of assets in smart contracts to prevent an adversary from tampering with or disrupting these processes. This countermeasure addresses threat T6. 
    \item \textbf{Embedding provenance updates in blockchain transactions.} This countermeasure leverages digital signatures and the inclusion of transactions into a transaction block to mitigate the spoofing and repudiation of provenance updates. This countermeasure addresses threat T3 and T4. 
\end{itemize}

\section{Discussions}

\subsection{Usage Scenarios}

\noindent\textbf{Auditing ML Assets:} The complete and trustworthy provenance information provided by ProML can help auditors detect issues of incoming ML assets more effectively. For instance, auditors at the organisation A in the running example can calculate and compare the cryptographic hash of the received ML model against the one reported by the developers at C to detect the model swapping attack by an insider threat at C. By inspecting the training and testing datasets reported by B and C via ProML, auditors at A can detect fairness risks of the received model early. Even if an organisation omits provenance updates from ProML, the omission itself can serve as an indicator of compromise. 

\vspace{1.5mm}
\noindent\textbf{Track and trace vulnerabilities and compromises:} Having a complete and trustworthy history helps auditors trace compromised assets to their root causes. For instance, if A determines that the model from C does not match the one reported by the developers at C, then both A and C can trace the compromised model to the employee who carries out the model delivery. After detecting the root cause (i.e., employee-turn-rogue), C can use ProML to track all activities that involve this employee to identify all potential compromises for damage control. 

\subsection{Where ProML performs unsatisfactorily}


\noindent\textbf{Case 1: Network fails to reach a threshold size.} The security of a decentralised system like ProML depends on the number of participants. Whilst existing surveys have suggested the existence of large-scale ML workflows, some ML workflows might not clear the threshold number of participants. In the future, we plan to explore mechanisms to combine ProML networks to pool their resources for processing blockchain transactions whilst maintaining the confidentiality of the constituting ProML networks. 

\vspace{1.5mm}
\noindent\textbf{Case 2: Participants deliberately submit misinformation.} Blockchain and smart contracts cannot detect misleading information or omission in the provenance information submitted by participants. However, they maintain trustworthy records of the submitted information that allow other participants to audit and detect misbehaviours. Future research could investigate blockchain-based peer-review protocols to leverage participants' expertise in auditing ML assets and their provenance.

\subsection{Threat to Validity}

The generalizability of the benchmark results is a validity threat to our evaluation. We mitigated this by choosing Ropsten, the most realistic representation of the Ethereum main net\footnote{\url{https://ethereum.org/en/developers/docs/networks/#ropsten}} which is one of the most prominent blockchains. Moreover, the performance of ProML would be higher in practice as organisations are likely to adopt consortium blockchains such as Hyperledger Fabric, which exchange openness and anonymity for orders of magnitude performance gain. Therefore, we believe that our benchmark results are helpful because they represent a worst-case scenario.

\section{Related Work}
\label{sec:related_work}

This paper aligns with the research on the secure management of the provenance of ML models and datasets. Earlier works on data provenance focused on logging the utilisation of computer files for future privacy compliance audits \cite{Neisse:2017:1,Sutton:2017:1} or mitigating poisoning attacks on ML models via datasets \cite{Baracaldo:2017:1,Stokes:2021:1}. Gebru et al. \cite{Gebru2021} were the first to present a standardized process via question answering to document datasets used in ML projects. Sutton et al. \cite{Sutton:2017:1} were among the firsts to leverage the Bitcoin blockchain as a security countermeasure to protect the log itself. Since then, blockchain has been and is still commonly used as an alternative secure storage solution for log data (e.g., \cite{Kumar:2018:1,Zhang:2018:1}). Recent research has extended blockchain's role by leveraging its programmability via smart contracts to extend its role. For instance, ProvHL \cite{Demichev:2018:1}, \cite{Dang:2020:1} and \cite{Khatal:2021:1} employed blockchain as an authoriser and logger for the data stored on an open-source data management system or an IPFS content network. Moller et al. \cite{Moller:2021:1} was among the first that captured workflow steps rather than access logs of electronic data on the blockchain as provenance information. 

The provenance of ML models has been relatively less explored. Earlier works on model provenance (e.g., \cite{Schelter:2017:1}) focused on the automated extraction of metadata and provenance of ML experiments, such as details of their training runs, statistics of datasets, evaluation results, and models' metadata. Mitchell et al. \cite{Mitchell2019} proposed a standardized document called model cards to accompanied trained models, presenting their performance characteristics and other metadata. Souza et al. \cite{Souza:2020:1} were among the first to advocate for describing the capturing the workflow provenance of ML models to address the increasingly complicated and heterogeneous nature of ML workflows. Recent years have witnessed increased utilisation of blockchain as a security mechanism for the provenance of ML models. For instance, Lo et al. \cite{lo2022towards} employed smart contracts to store the encrypted hashes of datasets and models appearing in federated learning systems. ProML does more than storing the hashes of models or datasets emerging from an ML workflow. Inspired by \cite{Souza:2020:1} and \cite{Moller:2021:1}, ProML also captures details regarding how workflow activities are performed on ML assets, empowering deeper auditing well as track and trace of compromises. We codified a structured way to capture workflow-centric provenance information on a blockchain with the proposed ASM approach. Moreover, instead of shifting the responsibility for managing provenance from a cloud service to a remote blockchain network, ProML fully embraces decentralisation, allowing relevant participants to secure the provenance information themselves and maintain complete control over how and what information they share.

\section{Conclusion}
\label{sec:conclusion}

Large-scale Machine Learning based Software Systems are increasingly developed by distributed ML workflows, where provenance information is paramount to secure and verify the circulated models and datasets. This paper proposes a decentralised platform named ProML that leverages blockchain and smart contracts for disseminating, storing, and updating ML provenance securely. ProML proposes a novel architectural approach called Artefact-as-a-State-Machine (ASM) to store and manage ML provenance on a blockchain. Via benchmarks conducted on a global test blockchain network, we showed that ProML's latencies are negligible compared to the average timeframe of ML workflows. Via a security analysis, we also showed that ProML is secure against tampering, spoofing, repudiation, and denial of service threats in a distributed ML workflow environment. Thus, we believe that ProML could be a foundation for developing decentralised software systems that help secure the increasingly remote and distributed engineering process of ML and software systems.

\bibliographystyle{splncs04}
\bibliography{references}

\end{document}